\documentclass[final,5p,times,twocolumn]{elsarticle}

\usepackage{amssymb}

\journal{J. Magn. Magn. Mater.}

\begin{document}

\begin{frontmatter}



\title{Magnetic properties, electrical resistivity, and hardness of high-entropy alloys FeCoNiPd and FeCoNiPt}


\author{Jiro Kitagawa$^1$}

\address{$^1$ Department of Electrical Engineering, Faculty of Engineering, Fukuoka Institute of Technology,3-30-1 Wajiro-higashi, Higashi-ku, Fukuoka 811-0295, Japan}
\ead{j-kitagawa@fit.ac.jp}

\begin{abstract}
We report the magnetic properties, electrical resistivity, and Vickers microhardness of as-cast and annealed high-entropy alloys (HEAs) FeCoNiPd and FeCoNiPt with the face-centered cubic structure. The heat treatment at 800 $^{\circ}$C does not largely affect the physical properties in each HEA. The values of the Curie temperature and the saturation moment at 50 K are 955 K and 1.458 $\mu_\mathrm{B}$/f.u. for the annealed FeCoNiPd, and 851 K and 1.456 $\mu_\mathrm{B}$/f.u. for the annealed FeCoNiPt, respectively. Each HEA is a soft ferromagnet and shows metallic resistivity. The electronic structure calculations of both HEAs support the ferromagnetic ground states. The comparisons between experimental and theoretical values are made for the Curie temperature, the saturation moment, and the residual resistivity. The Vickers microhardness of annealed FeCoNiPd and FeCoNiPt are both 188 HV. The hardness vs. valence electron count (VEC) per atom plot of these HEAs does not largely deviate from an expected universal relation forming a broad peak at VEC$\sim$6.8. This study would give some hints on designing a soft ferromagnetic HEA with high hardness.
\end{abstract}




\end{frontmatter}


\section{Introduction}
High-entropy alloys (HEAs) receive much interest due to their rich functionality, such as high strength, corrosion resistance, energy storage, radiation protection, superconductivity, soft ferromagnetism, and biocompatibility\cite{Sathiyamoorthi:PMS2022,Chang:AM2020,Wang:JMCA2021,Marques:EES2021,Pickering:Entropy2021,Kitagawa:Metals2020,Chaudhary:MT2021,Castro:Metals2021}.
The entropy state of an alloy is classified by the configurational entropy $\Delta S_\mathrm{mix}=-R\sum_{i=1}^{n}c_{i}\mathrm{ln}c_{i}$, where $n$ is the number of elements, $c_{i}$ is the atomic fraction, and $R$ is the gas constant.
At the initial stage of HEAs research, HEA required $\Delta S_\mathrm{mix}$ larger than 1.62 $R$, which can be realized by the solid solution of more than five elements.
The critical value defining HEA is now\cite{Yan:MMTA2021} 1.0 $R$.
So the equiatomic quaternary alloy investigated in this study can be referred to as HEA.
In HEAs, the microstructure often affects the physical properties, and, for example, the control of microstructure towards improved mechanical or magnetic properties is the central topic of HEAs\cite{Bhardwaj:TI2021,Rao:AFM2021}.
The other current topic is the materials research on HEAs exhibiting novel phenomena \cite{Zherebtsov:Int2020,Kitagawa:APLMater2022} by utilizing the large compositional space of HEAs.

Magnetic HEAs are attracting much attention because of their good soft ferromagnetism coexisted with high strength and/or high hardness\cite{Chaudhary:MT2021,Shivam:JALCOM2020}.
The magnetic HEAs are a promising alternative to conventional soft magnetic materials with poor strength\cite{Chaudhary:MT2021}.
Based on a face-centered cubic (fcc) FeCoNi with high saturation magnetization and low coercive field $H_\mathrm{c}$, many studies on the effect of adding alloying elements are carried out.
For example, Al$_{x}$CoCrFeNi is well investigated HEA\cite{Kao:JALCOM2011}, which shows a structural change from fcc to body-centered cubic (bcc) structure with increasing $x$.
The Curie temperature $T_\mathrm{C}$ of Al-free CoCrFeNi is 120 K, which can be enhanced above room temperature as $x$ increases\cite{Kao:JALCOM2011}.
NiFeCoCrPd and NiFeCoCrMn are famous equiatomic HEAs with $T_\mathrm{C}$=440 K and 38 K, respectively\cite{Billington:PRB2020,Schneeweiss:PRB2017}.
There are a few reports which aim at attaining a higher $H_\mathrm{c}$.
FeCoNiAlCu$_{x}$Ti$_{x}$ shows a relatively high $H_\mathrm{c}$ of 955 Oe after heat treatment\cite{Na:AIPAdv2021}.
The recent research on magnetic HEAs shows a growing interest in controlling magnetic properties through tailoring a microstructure\cite{Zuo:ActaMat2017,Rao:AFM2021}.
For example, Fe$_{15}$Co$_{15}$Ni$_{20}$Mn$_{20}$Cu$_{30}$ exhibits a spinodal decomposition after a heat treatment\cite{Rao:AFM2021}, which leads to enhanced $T_\mathrm{C}$.

\begin{table*}
\centering
\caption{\label{tab:table1}Lattice parameter $a$, atomic composition determined by EDX measurement, Curie temperature $T_\mathrm{C}$, saturation moment $M_\mathrm{s}$ at 50 K, and residual electrical resistivity $\rho_\mathrm{res}$ of each sample.}
\begin{tabular}{cccccc}
\hline
 sample & a (\AA) & atomic composition & $T_\mathrm{C}$ (K) & $M_\mathrm{s}$ ($\mu_\mathrm{B}$/f.u.) & $\rho_\mathrm{res}$ ($\mu\Omega$cm) \\
\hline
FeCoNiPd (as-cast) & 3.683(2) & Fe$_{24.8(8)}$Co$_{26(1)}$Ni$_{25.2(5)}$Pd$_{24(1)}$ & 947 & 1.478 & 9.6    \\
FeCoNiPd (annealed) & 3.681(1) & Fe$_{25.2(9)}$Co$_{24(1)}$Ni$_{25.5(5)}$Pd$_{25.3(4)}$ & 955 & 1.458 & 8.1    \\
FeCoNiPt (as-cast) & 3.692(1) & Fe$_{25.4(6)}$Co$_{24(1)}$Ni$_{23.8(6)}$Pt$_{26.8(9)}$ & 848 & 1.464 & 26.1   \\
FeCoNiPt (annealed) & 3.686(1) & Fe$_{24(1)}$Co$_{25.0(5)}$Ni$_{26.0(7)}$Pt$_{25.0(7)}$ & 851 & 1.456 & 21.2    \\
\hline
\end{tabular}
\end{table*}

We are focusing on HEAs with the combination of 3$d$ magnetic elements and noble metals (e.g., Rh, Ir, Pd, and Pt)\cite{Baba:Materials2021} because materials research on such an HEA is unexplored.
Recently, Fukushima et al. reported the database ($T_\mathrm{C}$, spin moment, residual resistivity) of 147630 quaternary HEAs produced by a density functional theory calculation\cite{Fukushima:PRM2022}.
This database contains fcc FeCoNiPd and fcc FeCoNiPt, which can be regarded as Pd- or Pt-added FeCoNi alloy, and the magnetic properties of these HEAs are not well investigated.
Only the M\"{o}ssbauer effect of FeCoNiPd is reported\cite{Ciealak:JMMM2021}, and the fundamental magnetic properties such as $T_\mathrm{C}$ and saturation moment $M_\mathrm{s}$ are unknown.
Therefore, this study's first purpose is to assess the fundamental magnetic properties of fcc FeCoNiPd and fcc FeCoNiPt and to compare the magnetic properties between these HEAs and FeCoNi.

The Vickers microhardness is usually measured to evaluate the hardness of a material.
In HEAs, it is reported that the Vickers microhardness empirically correlates with the valence electron count (VEC) per atom\cite{Tian:IM2015}.
The hardness vs. VEC plot of HEAs with VEC ranging from 4.1 to 8.8 seems to form a broad peak at VEC$\sim$6.8, as mentioned below.
The VEC of FeCoNiPd or FeCoNiPt is 9.25, and we are interested in the hardness to check the possible universal relationship between the hardness and the VEC.
This examination would be useful for designing soft ferromagnetic HEAs with high hardness.
The second purpose of this study is to investigate the Vickers microhardness of FeCoNiPd and FeCoNiPt.

This paper reports the magnetic properties, electrical resistivity, and hardness of as-cast and annealed fcc FeCoNiPd and fcc FeCoNiPt.
The electronic structure was calculated to elucidate the ferromagnetism in each HEA.
We have found that the heat treatment does not significantly influence the physical properties of FeCoNiPd and FeCoNiPt.
The soft ferromagnetic behaviors are observed in both HEAs.
The comparisons between experimental and theoretical values are made for $T_\mathrm{C}$, $M_\mathrm{s}$, and the residual resistivity.
The hardness of FeCoNiPd and FeCoNiPt do not deviate from the empirical relationship between the hardness and the VEC.

\section{Materials and Methods}
Polycrystalline samples of FeCoNiPd and FeCoNiPt were synthesized by a home-made arc furnace using constituent elements Fe (99.9 \%), Co (99.9 \%), Ni (99.9 \%), Pd (99.9 \%), and Pt (99.9 \%) under Ar atmosphere.
The button-shaped samples were remelted several times on a water-cooled Cu hearth and flipped each time to ensure homogeneity.
The as-cast samples were annealed in an evacuated quartz tube at 800 $^{\circ}$C for four days.
Room temperature X-ray diffraction (XRD) patterns of samples were recorded using an X-ray diffractometer (XRD-7000L, Shimadzu) with Cu-K$\alpha$ radiation.
We used thin slabs cut from the samples due to their high ductility.
The scanning electron microscope (SEM) images were collected using a field emission scanning electron microscope (FE-SEM, JSM-7100F, JEOL).
The chemical composition was also evaluated by an energy dispersive X-ray (EDX) spectrometer equipped with the FE-SEM.

The temperature dependence of dc magnetization $\chi_\mathrm{dc}$ ($T$) between 50 and 300 K was measured using VersaLab (Quantum Design).
The high-temperature $\chi_\mathrm{dc}$ ($T$) from 300 K to 1173 K was measured by a vibrating sample magnetometer (TM-VSM33483-HGC, Tamakawa) to estimate $T_\mathrm{C}$.
The isothermal magnetization ($M$) curve at 50 K was taken using the VersaLab.
The temperature dependence of electrical resistivity $\rho$ ($T$) between 3 K and 300 K was measured by a conventional dc four-probe method using a home-made sysytem in a GM refrigerator (UW404, Ulvac cryogenics).
The Vickers microhardness was measured under the applied load of 0.49, 0.98, 1.96, 2.94, and 4.903 N, respectively, using a Shimadzu HMV-2T microhardness tester.
The holding time under the diamond indenter is 10 s.

We also performed the electronic structure calculation using the coherent potential approximation (CPA) approach because of no report of the density of states (DOS) in the previous study\cite{Fukushima:PRM2022}.
We employed the Akai-KKR program package\cite{Akai:JPSJ1982}, which is based on the Korringa-Kohn-Rostoker (KKR) method with CPA.
We used the generalized gradient approximation from Perdew-Burke-Ernzerhof (PBE) and treated the spin-polarization and the spin-orbit interaction.
The spin-orbit interaction is not included in the previous study\cite{Fukushima:PRM2022}.

\section{Results and Discussion}

\begin{figure}
\centering
\includegraphics[width=0.9\linewidth]{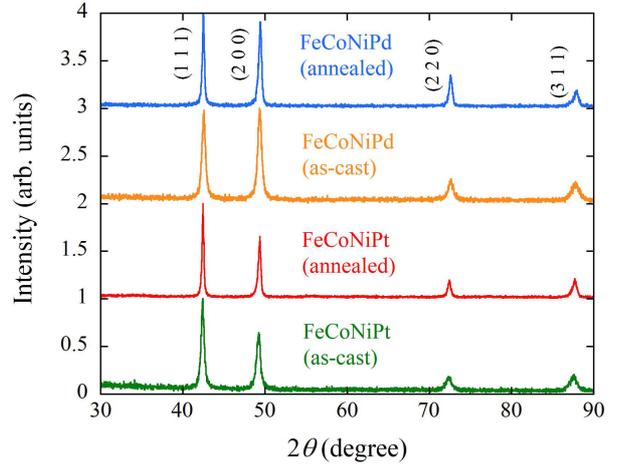}
\caption{\label{fig1} XRD patterns of FeCoNiPd and FeCoNiPt. The origin of each pattern is shifted by a value for clarity.}
\end{figure}

\begin{figure*}
\centering
\includegraphics[width=0.8\linewidth]{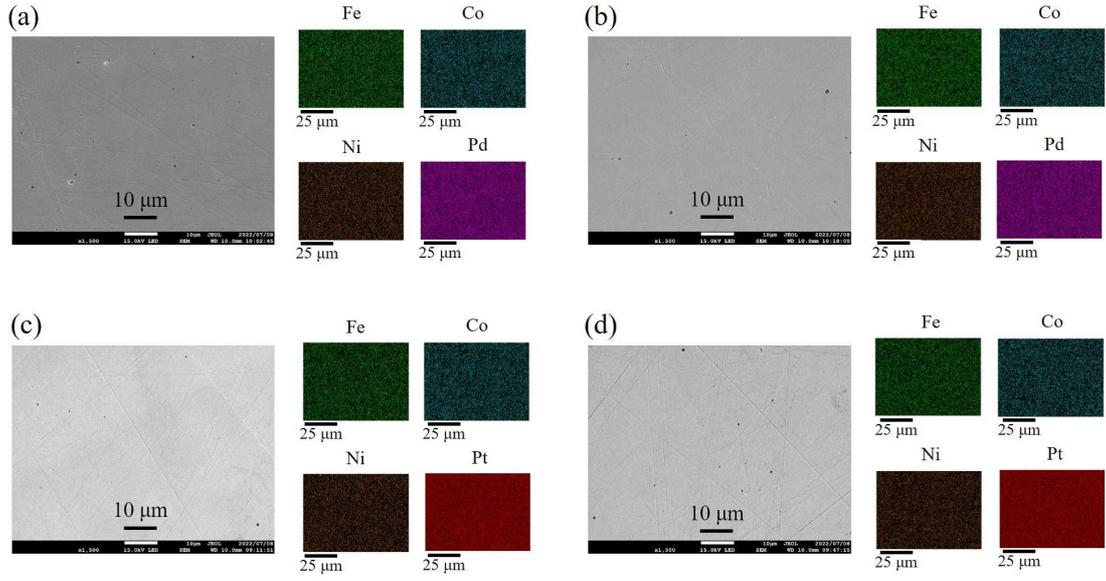}
\caption{\label{fig2} SEM images of (a) FeCoNiPd (as-cast), (b) FeCoNiPd (annealed), (c) FeCoNiPt (as-cast), and (d) FeCoNiPt (annealed), respectively. The elemental mappings are also shown.}
\end{figure*}

\begin{figure*}
\centering
\includegraphics[width=0.85\linewidth]{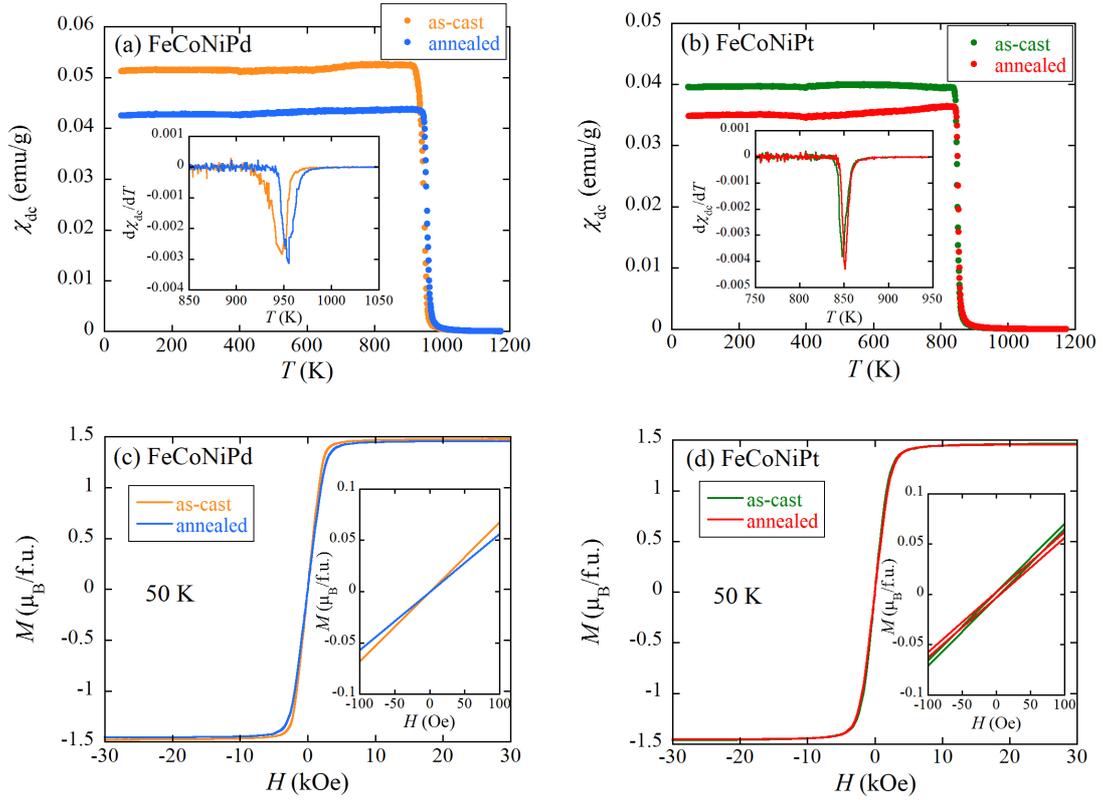}
\caption{\label{fig3} (a) Temperature dependences of $\chi_\mathrm{dc}$ of as-cast and annealed FeCoNiPd. The external field is 100 Oe. The inset is the temperature derivative of $\chi_\mathrm{dc}$ for each sample. (b) Temperature dependences of $\chi_\mathrm{dc}$ of as-cast and annealed FeCoNiPt. The external field is 100 Oe. The inset is the temperature derivative of $\chi_\mathrm{dc}$ for each sample. (c) Isothermal magnetization curves at 50 K of as-cast and annealed FeCoNiPd. The inset is the expanded view. (d) Isothermal magnetization curves at 50 K of as-cast and annealed FeCoNiPt. The inset is the expanded view.}
\end{figure*}

Figure \ref{fig1} shows the XRD patterns of as-cast and annealed FeCoNiPd and FeCoNiPt.
All patterns can be indexed by fcc structure with the Miller indices denoted in the figure.
The lattice parameters $a$ obtained by the least square method are listed in Table \ref{tab:table1}.
SEM images and elemental mappings of all samples are displayed in Fig.\ref{fig2}.
In each sample, no trace of impurity phase is detected, and the elemental mapping shows a homogeneous elemental distribution.
The chemical compositions evaluated by EDX measurements are tabulated in Table \ref{tab:table1} and agree well with the ideal one with 25 at.\% for each element.
HEAs often show composition segregation after heat treatment\cite{Vrtnik:JALCOM2017,Pacheco:InorgChem2019}.
However, FeCoNiPd or FeCoNiPt forms a stable single-phase fcc against the heat treatment at 800 $^{\circ}$C.

Figures \ref{fig3}(a) and (b) show $\chi_\mathrm{dc}$ ($T$) under the external field $H$ of 100 Oe for FeCoNiPd and FeCoNiPt, respectively.
In each sample, a steep increase of $\chi_\mathrm{dc}$ is observed as the temperature is lowered, which indicates a ferromagnetic ordering.
$T_\mathrm{C}$ is estimated by the temperature derivative of $\chi_\mathrm{dc}$ and defined by the minimum point described in the inset of Fig.\ref{fig3}(a) or (b).
This is one of the effective ways to obtain $T_\mathrm{C}$ in transition metal-based ferromagnets\cite{Oikawa:APL2001,Yu:APL2003,Kitagawa:JMMM2018,Kitagawa:JSSC2020}.
Thus obtained $T_\mathrm{C}$s are listed in Table \ref{tab:table1}, and the slight enhancement of $T_\mathrm{C}$ after the annealing is confirmed in each HEA.
Fukushima et al. have provided $T_\mathrm{C}$ data obtained by the density functional theory calculation for 147630 quaternary HEAs and placed the data in a repository\cite{Fukushima:rep}.
The predicted $T_\mathrm{C}$s of fcc FeCoNiPd and fcc FeCoNiPt are 1137 and 1085 K, respectively.
Although the higher $T_\mathrm{C}$ of FeCoNiPd compared to FeCoNiPt is consistent with the experimental result, the theoretical value is approximately 200 K higher than the experimental one in each HEA.
The mean-field approximation used for calculating $T_\mathrm{C}$ tends to overestimation, which is the reason for the relatively large difference between theoretical and experimental values.
The isothermal $M$-$H$ curves of FeCoNiPd and FeCoNiPt measured at 50 K are exhibited in Figs.\ref{fig3}(c) and (d).
With the increase of $H$ from 0 Oe in each HEA, $M$ increases steeply and soon saturates, which supports the ferromagnetic ground state.
Thermal annealing seems not to affect $M$-$H$ curves.
The values of $M_\mathrm{s}$ for HEAs investigated are summarized in Table \ref{tab:table1} and compared with the theoretical values below.
The inset of each figure is the expanded view to show the hysteresis.
The very weak hysteresis indicates the soft ferromagnetism in each HEA.
While $H_\mathrm{c}$ of FeCoNiPt would be approximately 2 Oe, no hysteresis is observed for FeCoNiPd within the measurement accuracy.

\begin{figure}
\centering
\includegraphics[width=0.7\linewidth]{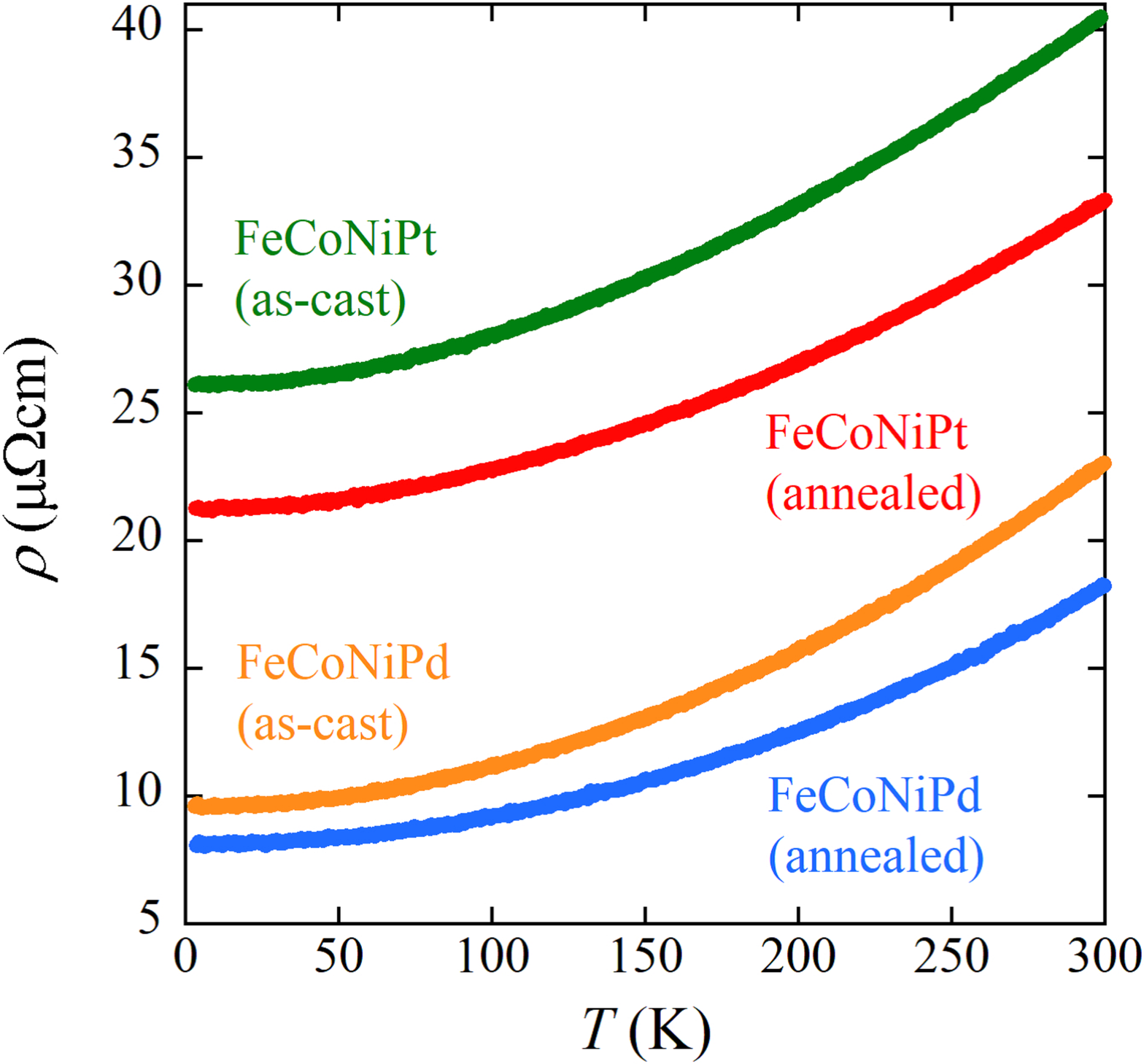}
\caption{\label{fig4} Temperature dependences of $\rho$ of FeCoNiPd and FeCoNiPt.}
\end{figure}

\begin{figure*}
\centering
\includegraphics[width=0.8\linewidth]{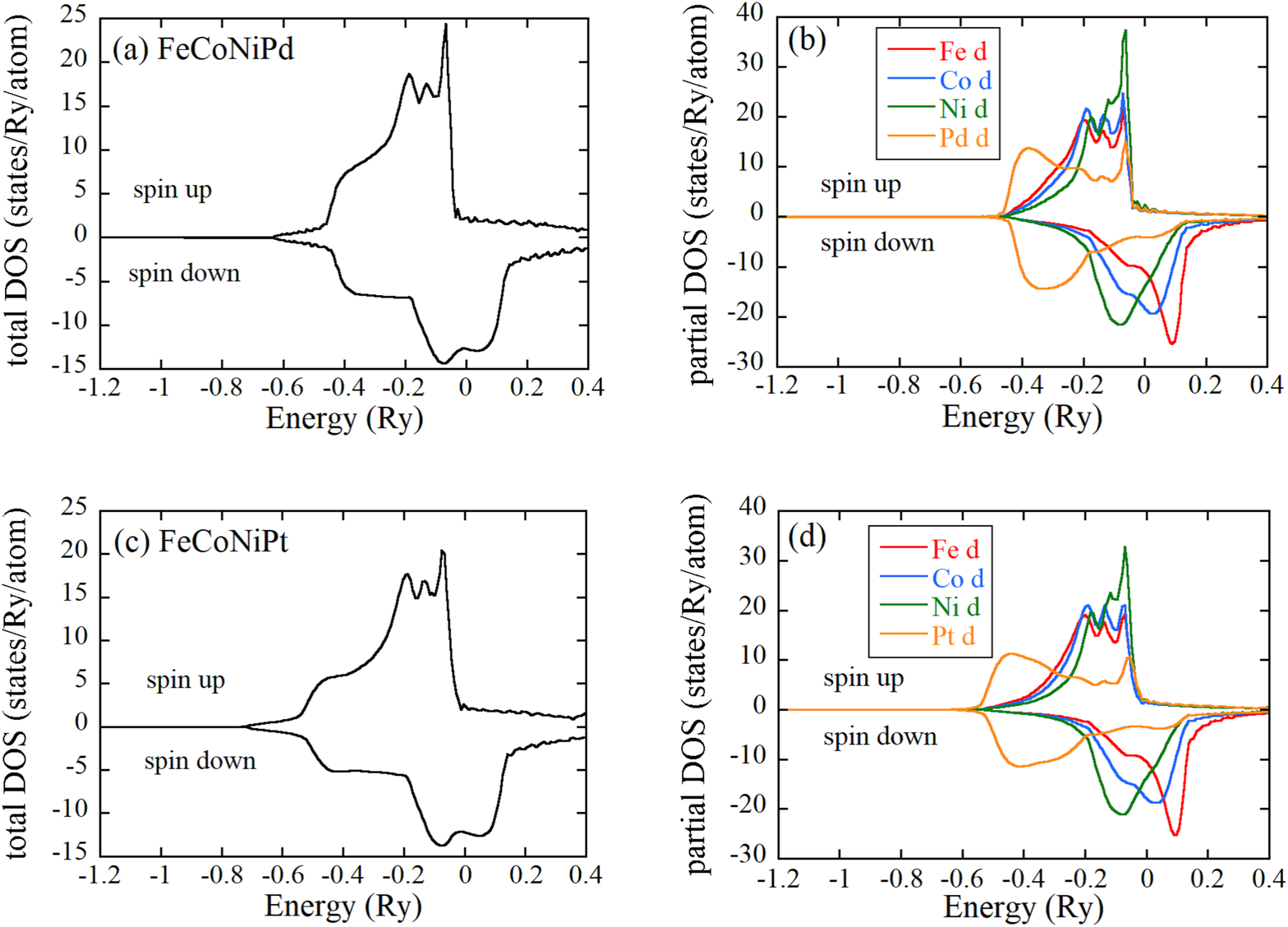}
\caption{\label{fig5} Electronic density of states of fcc FeCoNiPd ((a) and (b)) and fcc FeCoNiPt ((c) and (d)). Each partial DOS ((b) and (d)) is drawn for only $d$-electrons due to the dominant contribution around the Fermi level. The Fermi energy is set to 0 Ry.}
\end{figure*}

Here we compare the magnetic properties of FeCoNiPd and FeCoNiPt with those of fcc FeCoNi.
The lattice parameter, $T_\mathrm{C}$, and $M_\mathrm{s}$ of fcc FeCoNi are reported to be 3.599 \AA, 1000 K, and 163 emu/g (=1.687 $\mu_\mathrm{B}$/f.u.), respectively\cite{Zuo:JMMM2014,Lin:MCP2021}.
The atomic radius of Pd(Pt) is 1.3754 \AA (1.387 \AA), which is larger than those of Fe, Co, and Ni (1.2412, 1.2510, and 1.2459 \AA, respectively)\cite{Miracle:ActMat2017}.
Therefore, the lattice parameter of FeCoNiPd or FeCoNiPt expands compared to FeCoNi.
As discussed later, Pd(Pt) carries the magnetic moment smaller than Fe, Co, and Ni moments, which causes the reduction of $M_\mathrm{s}$ in FeCoNiPd or FeCoNiPt.
Going from FeCoNi, FeCoNiPd to FeCoNiPt, the lattice parameter increases, and $T_\mathrm{C}$ is systematically reduced.
The expansion of lattice means the increase of interatomic distance between 3$d$ elements, which would lead to a weakened magnetic exchange interaction.
In this case, the systematic reduction of $T_\mathrm{C}$ with increasing lattice parameter can be anticipated.
We note that FeCoNiCr ($a$=3.580 \AA, $T_\mathrm{C}$=104 K)\cite{Chou:MSEB2009,Na:AIPAdv2018} and FeCoNiMn ($a$=3.6029 \AA, $T_\mathrm{C}$=332 K)\cite{Rao:PRM2020} significantly reduce $T_\mathrm{C}$, which is independent of the unit-cell volume change compared to FeCoNi.
The alloying by antiferromagnetic elements of Cr or Mn severely decreases the averaged magnetic exchange interaction strength.
FeCoNi possesses $H_\mathrm{c}$ of 1.5 Oe and this value is not largely enhanced in fcc FeCoNiAl$_{x}$ or fcc FeCoNiSi$_{x}$ ($H_\mathrm{c}$: 0.5$\sim$6 Oe)\cite{Chaudhary:MT2021}.
The same trend is observed in FeCoNiPd and FeCoNiPt, and the soft ferromagnetism of fcc FeCoNi is robust against the addition of alloying elements.

Next, $M_\mathrm{s}$ and $H_\mathrm{c}$ of FeCoNiPd and FeCoNiPt are compared with those of related systems.
The values of $M_\mathrm{s}$ in the unit of emu/g for FeCoNiPd and FeCoNiPt after annealing are 116 emu/g and 88 emu/g, respectively.
The corresponding values of FeCoNi-related systems\cite{Deng:MC2020,Dasari:Mat2020,Zuo:JMMM2014,Li:IM2017} are, for example, 130 emu/g in FeCoNiAl$_{x}$ ($x$=0.2 and 0.3), 126 emu/g in FeCoNiSi$_{0.25}$, and 101 emu/g in FeCoNiAl$_{0.25}$Mn$_{0.25}$.
So, while $M_\mathrm{s}$ of FeCoNiPd is slightly reduced compared to FeCoNiAl$_{x}$ and FeCoNiSi$_{0.25}$, the relatively large reduction of $M_\mathrm{s}$ occurs in FeCoNiPt due to the heavy element Pt with a small magnetic moment.
This comparison indicates that light elements are favorable for achieving high $M_\mathrm{s}$ soft ferromagnet.
FeCoNiAl$_{x}$ ($x$=0.2 and 0.3), FeCoNiSi$_{0.25}$, and FeCoNiAl$_{0.25}$Mn$_{0.25}$ possess $H_\mathrm{c}$ ranging from 0.5 to 5 Oe\cite{Deng:MC2020,Dasari:Mat2020,Zuo:JMMM2014,Li:IM2017}.
These values are comparable to $H_\mathrm{c}$ ($\sim$2 Oe) of FeCoNiPt.
On the other hand, $H_\mathrm{c}$ of FeCoNiPd would be smaller compared to the related systems.
$H_\mathrm{c}$ of HEAs and commercial soft ferromagnets are summarized in the review by Huang et al\cite{Huang:crystals2020}.
According to the review, $H_\mathrm{c}$ of FeCoNiPd would locate in the range of commercial soft ferromagnets (Ni-Fe alloys with $H_\mathrm{c}$ of 0.004 Oe $\sim$ 0.1 Oe).

Figure \ref{fig4} summarizes $\rho$ ($T$) of as-cast and annealed FeCoNiPd and FeCoNiPt.
For each HEA, $\rho$ of the as-cast sample decreases after the heat treatment, which suggests a relaxation of the lattice distortion.
Despite the existence of atomic disorders, $\rho$ smoothly decreases with cooling.
The measured temperature range is well below $T_\mathrm{C}$, and relatively large temperature dependences mean the dominance of magnetic contribution to the electrical transport.
Such a behavior is often observed in ferromagnetic metals with atomic disorders\cite{Kitagawa:Metals2020-2}.
The database\cite{Fukushima:rep} by Fukushima et al. also includes the values of residual resistivity $\rho_\mathrm{res}$.
According to the database, $\rho_\mathrm{res}$s of fcc FeCoNiPd and fcc FeCoNiPt are 5.1 and 21.6 $\mu\Omega$cm, respectively, which agree well with the experimental values of annealed HEAs.

\begin{table*}
\centering
\caption{\label{tab:table2}Spin moment and orbital moment of each element in FeCoNiPd and FeCoNiPt with fcc structure obtained by electronic structure calculation. The total moments are 1.402 $\mu_\mathrm{B}$/f.u. and 1.421 $\mu_\mathrm{B}$/f.u. for FeCoNiPd and FeCoNiPt, respectively.}
\begin{tabular}{cccccc}
\hline
 \multicolumn{3}{c}{FeCoNiPd}&\multicolumn{3}{c}{FeCoNiPt}\\
atom & spin moment ($\mu_\mathrm{B}$) & orbital moment ($\mu_\mathrm{B}$) & atom & spin moment ($\mu_\mathrm{B}$) & orbital moment ($\mu_\mathrm{B}$)\\
\hline
Fe & 2.854 & 0.0661 & Fe & 2.875 & 0.0657 \\
Co & 1.874 & 0.1015 & Co & 1.902 & 0.0891 \\
Ni & 0.779 & 0.0641 & Ni & 0.806 & 0.0505 \\
Pd & 0.233 & 0.0139 & Pt & 0.242 & 0.0413 \\
\hline
\end{tabular}
\end{table*}

In the electronic structure calculations of fcc FeCoNiPd and fcc FeCoNiPt, the lattice parameters obtained for the annealed samples are used, and the perfect solid solution of constituent elements is assumed.
Figures \ref{fig5} (a) and (c) exhibit the total DOSs of FeCoNiPd and FeCoNiPt, respectively.
In each case, the difference in total DOS between spin-up and spin-down electrons supports the ferromagnetic ground state.
Partial DOSs are displayed in Figs.\ref{fig5} (b) and (d) for FeCoNiPd and FeCoNiPt, respectively.
Only $d$-electron DOS is drawn for each partial DOS due to the dominant contribution around the Fermi level.
Reflecting the isoelectronic HEAs, the structures of partial DOS shown in Figs.\ref{fig5} (b) and (d) for each element are similar to each other.
The partial DOSs indicate the presence of magnetic moment for all elements.
The spin and orbital moment values calculated for all elements are summarized in Table \ref{tab:table2}.
In each HEA, all moments align parallel, and Fe and Co spin moments are dominant.
The spin moment values of Fe, Co, and Ni are slightly larger than those obtained by Fukushima et al\cite{Fukushima:rep}.
Accordingly, the total moments of 1.402 $\mu_\mathrm{B}$/f.u. for FeCoNiPd and 1.421 $\mu_\mathrm{B}$/f.u. for FeCoNiPt are also respectively larger than 1.383 and 1.395 $\mu_\mathrm{B}$/f.u. in the database.
The spin-orbit interaction considered in this study would be responsible for the slight increase of moment.
The calculated total moments in this study or the database well explain the experimental $M_\mathrm{s}$s (see also Table \ref{tab:table1}), which means that the magnetic structure of FeCoNiPd or FeCoNiPt would be a simple one with all spins aligning parallel.

\begin{figure}
\centering
\includegraphics[width=0.75\linewidth]{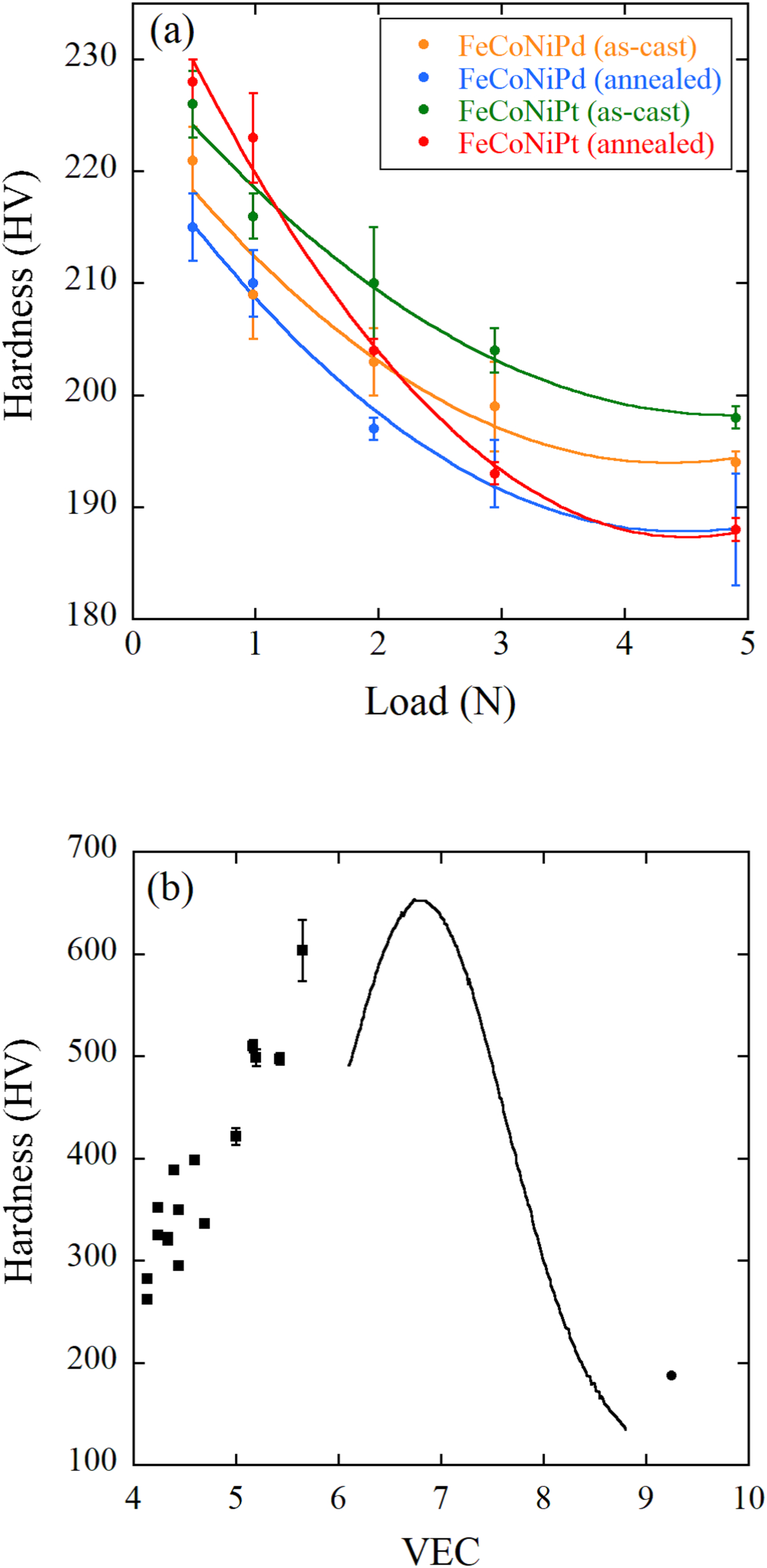}
\caption{\label{fig6} (a) Load dependences of Vickers microhardness for FeCoNiPd and FeCoNiPt. The solid curves are guides for the eyes. (b) VEC dependence of Vickers microhardness of FeCoNiPd and FeCoNiPt with VEC=9.25. The refractory bcc HEAs reported in the literatures\cite{Kitagawa:JALCOM2022,Han:IM2017,Li:Materials2019,Bhandari:JMRT2020,Ge:MSEA2020} are also plotted (VEC:4.14$\sim$5.65). The solid curve represents the guideline obtained for HEAs with bcc structures (VEC: 6.0$\sim$7.55) and fcc structures (VEC: 7.8$\sim$8.8) taken from the literature\cite{Tian:IM2015}.}
\end{figure}

Figure \ref{fig6} (a) depicts the Vickers microhardness vs. applied load profiles for all samples.
The hardness in each sample gradually decreases as the load is increased, which is also reported in many HEAs and intermetallic compounds\cite{Zhu:JALCOM:2022,Ma:JSSC2021}.
It is well known that the elastic recovery effect is responsible for the increase of Vickers microhardness with decreasing load\cite{Jindal:SC1988}.
In FeCoNiPd, the annealed sample displays lower hardness than the as-cast sample.
The annealing would cause the release of lattice distortion introduced in the rapid solidification process of arc melting, which leads to lower hardness.
FeCoNiPt also shows similar behavior.
It is proposed that the Vickers microhardness depends on the VEC.
Figure \ref{fig6}(b) presents the VEC dependence of hardness.
The solid curve is a guideline obtained for HEAs with the bcc structures (VEC: 6.0$\sim$7.55) and fcc structures (VEC: 7.8$\sim$8.8)\cite{Tian:IM2015}.
The result of refractory bcc HEAs with VEC:4.14$\sim$5.65 is also shown\cite{Kitagawa:JALCOM2022,Han:IM2017,Li:Materials2019,Bhandari:JMRT2020,Ge:MSEA2020} and seems to be connected to the guideline.
In addition, a deep learning study of the hardness of refractory HEAs with the VEC: 4$\sim$6 also supports the positive correlation between the hardness and VEC in that VEC range\cite{Bhandari:Crystals2021}.
The VEC of FeCoNiPd or FeCoNiPt is 9.25, and the hardness of 188 HV obtained at 4.9030 N load is employed because the hardness at a higher load is usually employed. In addition we cannot obtain the hardness under a load higher than 4.9030 N due to the limitation of the microhardness tester used.
Thus plotted data points in Figure \ref{fig6}(b) do not essentially deviate from the expected universal relation between the VEC and the hardness.
We note that there are not enough data points from other HEAs at VEC larger than 9. Therefore, further study is required to prove the universal relation.
This study implies a vital role of VEC in designing the hardness of magnetic HEAs.
The VEC is also related to the phase stability of fcc and bcc HEAs\cite{Gao:book}: a single bcc phase for VEC between 5.0 and 6.87 and a single fcc phase for VEC larger than 8.0.
Therefore, if we need a soft ferromagnetic HEA with high hardness within the limitation of a single fcc phase, a magnetic HEA based on FeCoNi with an alloying for tuning VEC=8 is desirable.

\section{Summary}
We have investigated the fundamental magnetic properties, electrical resistivity, and Vickers microhardness of as-cast and annealed fcc FeCoNiPd and fcc FeCoNiPt.
After the heat treatment at 800 $^{\circ}$C, both HEAs keep the single-phase fcc, and the annealing does not largely alter the physical properties.
The values of $T_\mathrm{C}$ and $M_\mathrm{s}$ at 50 K are 955 K and 1.458 $\mu_\mathrm{B}$/f.u. for annealed FeCoNiPd, and 851 K and 1.456 $\mu_\mathrm{B}$/f.u. for annealed FeCoNiPt, respectively.
The coercive fields of both HEAs are very small, and they are soft ferromagnets.
In each HEA, the electrical resistivity shows the metallic temperature dependence.
The electronic structure calculations of both HEAs were performed, and the ferromagnetic ground states were obtained.
The total moments are close to those reported in the database made by the density functional theory calculation and agree with the experimental $M_\mathrm{s}$s.
The theoretical values of $T_\mathrm{C}$ and $\rho_\mathrm{res}$ are also reported in the database.
While the theoretical $T_\mathrm{C}$ values are relatively higher than the experimental ones in both HEAs, good agreement is confirmed between the theoretical and experimental $\rho_\mathrm{res}$ values.
The database would be beneficial for the materials research on magnetic quarternary HEAs.
In addition, the comparisons of magnetic properties of HEAs investigated and FeCoNi are made.
The Vickers microhardness of FeCoNiPd or FeCoNiPt slightly decreases after the annealing.
The hardness vs. VEC plot of these HEAs does not essentially deviate from the expected universal relation forming a broad peak at VEC$\sim$6.8.
This study provides a possible material design of soft ferromagnetic HEA with high hardness; a magnetic HEA based on FeCoNi with an alloying for tuning VEC=8 is promising.

\section*{ACKNOWLEDGMENTS}
J.K. is grateful for the support provided by the Comprehensive Research Organization of Fukuoka Institute of Technology.

\end{document}